\documentclass{emulateapj}

\usepackage{hyperref}
\usepackage{enumitem}
\usepackage{natbib}

\slugcomment{Accepted on October 30, 2013}

\shorttitle{Optimized PCA on Coronagraphic Images of Fomalhaut}
\shortauthors{Meshkat et al.}

\begin{document}

\title{Optimized Principal Component Analysis on Coronagraphic Images of the Fomalhaut System}

\author{Tiffany Meshkat$^1$, Matthew Kenworthy$^1$, Sascha P. Quanz$^2$, Adam Amara$^2$}
\altaffiltext{1}{Sterrewacht Leiden, P.O. Box 9513, Niels Bohrweg 2, 2300 RA Leiden, The Netherlands}  
\altaffiltext{2}{Institute for Astronomy, ETH Zurich, Wolfgang-Pauli-Strasse 27, 8093 Zurich, Switzerland}

\begin{abstract}

We present the results of a study to optimize the principal component analysis (PCA) algorithm for planet detection, 
a new algorithm complementing ADI and LOCI for increasing the contrast achievable next to a bright star.
The stellar PSF is constructed by removing linear combinations of principal components, allowing the flux 
from an extrasolar planet to shine through. The number of principal components used determines how well the stellar PSF 
is globally modelled. Using more principal components may decrease the number of speckles in the final image, but 
also increases the background noise. We apply PCA to Fomalhaut VLT NaCo images acquired at 4.05 $\mu$m with an apodized 
phase plate. We do not detect any companions, with a model dependent upper mass limit of 13-18 M$_{\text{Jup}}$ from 4-10 AU. 
PCA achieves greater sensitivity than the LOCI algorithm for the Fomalhaut coronagraphic data by up to 1 magnitude. 
We make several adaptations to the PCA code and determine which of these prove the most 
effective at maximizing the signal-to-noise from a planet very close to its parent star. We demonstrate that optimizing 
the number of principal components used in PCA proves most effective for pulling out a 
planet signal.

\end{abstract}


\section{Introduction}

The detection and characterization of extrasolar planets has grown dramatically as a field since 
the first detection in 1992 \citep{Wolszczan92}. The most successful detection techniques thus far 
are radial velocity and transit detection. Using ground and space based surveys (HARPS, Kepler, COROT, etc), 
these indirect techniques have discovered over 800 planets (exoplanet.eu) as well as thousands 
more planet candidates.

The direct detection of planets provides a unique opportunity to study exoplanets in the context of their formation 
and evolution. It complements the underlying semi-major axis exoplanet distribution from RV surveys
(from 100 AU down to a few AUs) and enables the characterization of the planet itself with an examination of its emergent flux as a 
function of wavelength. 
The detection of the planets HR8799 bcde \citep{Marois08}, Fomalhaut b \citep{Kalas08}, $\beta$ Pic b \citep{Lagrange09}, 
2MASS1207 \citep{Chauvin04}, 1RXS J1609-2105 b \citep{Lafreniere08}, HD 95086 b \citep{Rameau13b}, KOI-94 \citep{Takahashi13}
as well as discoveries of protoplanetary candidates LkCa 15 b \citep{Kraus12} and HD100546 b \citep{Quanz13},
demonstrate the potential breakthroughs of the technique.
However, thus far, most dedicated high contrast imaging surveys 
have yielded null-results (e.g. \citet{Rameau13},\citet{Vigan12}, \citet{Chauvin10}, \citet{Biller07},\citet{Heinze08}). 
These null results are due to the lack of contrast at small orbital separations, where most planets are expected to 
be found. Since planets are concluded to be rare at large orbital separations (\citet{Chauvin10}, \citet{Lafreniere07}), 
high contrast imaging must probe close to the parent star to detect a planet. 

High contrast imaging is limited by the diffraction limit, set by the 
telescope optics, which determines the minimum angular separation achievable under ideal conditions.
Since planets are low mass, cold, and red compared to their parent star (\citet{Spiegel12},\citet{Baraffe03}), 
the contrast ratio of their magnitudes is an additional constraint on their detectability. 
New instruments and techniques have been developed to combat these constraints at the acquisition and image 
processing stage.

Coronagraphs have been developed to reduce the light scattered in the 
telescope optics from diffraction during acquisition, but at a cost of throughput and angular resolution \citep{Guyon05}. 
Coronagraphic optics allow us to probe smaller inner working angles, but are limited by 
the stellar ``speckles'' which can dominate the flux from a planet \citep{Hinkley09}.

By turning off the telescope derotator on an alt-az telescope, the planet is able to ``rotate'' around the star, while the stellar PSF 
stays relatively stable and the speckles vary randomly in time. This technique is used in angular differential imaging (ADI, \citet{Marois08}). 
It takes advantage of this rotation to identify and subtract (in post-processing) the contribution from the stellar PSF 
and speckles. There are a number of image 
processing techniques aimed at modelling and subtracting the stellar PSF from every image, allowing the sky fixed planet signal to shine through. 
LOCI \citep{Lafreniere07} is an extension of ADI, which models the local stellar PSF structure in every image. 
Principal component analysis (PCA) ( \citet{Amara12}, \citet{Soummer12}, \citet{Brandt12}) models how the PSF varies in time by identifying 
the main linear components of the variation. Application 
of these image processing techniques has been demonstrated to increase the limiting magnitude achievable by up to a factor of 5 
(\citet{Lafreniere07}, \citet{Amara12}).

In this work, we present a detailed study of LOCI and PCA image processing techniques in order to optimize the 
signal-to-noise of a planet at small angular separations with the APP coronagraph (\citet{Kenworthy10},\citet{Kenworthy07}). 
We compare our results to the previous result of \citet{Kenworthy13}.

The Fomalhaut dataset that is used in the following analyses are a deep but typical observing sequence and will
act as an example for the rest of our surveys. 

\section{Data}

Data were obtained of the star Fomalhaut at the VLT/UT4 with NaCo 
(\citet{Lenzen03}, \citet{Rousset03}) in July and August 2011 (087.C-0701(B))
and were analyzed and published in \citet{Kenworthy13}. 
Fomalhaut was used as the natural guide star 
with the visible band wavefront sensor. The L27 camera on NaCo was used 
with the NB4.05 filter ($\lambda$ = 4.051$\mu$m and $\Delta\lambda$ = 0.02$\mu$m) and the Apodizing 
Phase Plate coronagraph (APP, \citet{Kenworthy10}, \citet{Quanz10}) to provide additional diffraction 
suppression. We used pupil tracking mode 
to perform ADI \citep{Marois06}. The PSF core is intentionally saturated to increase the signal from any potential companions. 

\begin{figure}
 \centering
 \epsscale{1.1}
 \plotone{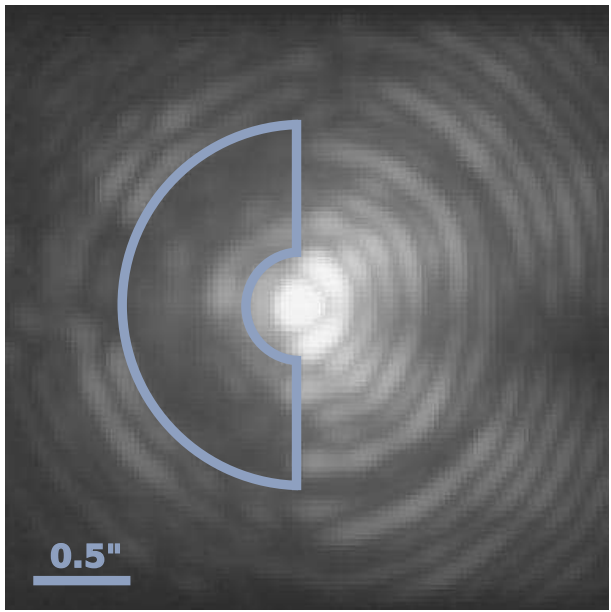}
 \caption{Image demonstrating the APP airy diffraction pattern with the diffraction suppressed region 
 outlined in blue. This is the only region that is used in the data reduction.}
 \label{app}
\end{figure}

The APP provides diffraction supression over a 180$^{\circ}$ wedge on one side of the target (\autoref{app}). Additional 
observations are required with a different position angle to cover the full 360$^{\circ}$ around the star. 
For these observations, we have three different datasets with different position angles ensuring 
full PA coverage around the target star. Each dataset has a large amount of field rotation: 119$^{\circ}$, 
117$^{\circ}$, 120$^{\circ}$.

Data were acquired in cube mode. Each data cube contains 200 frames, each with an integration time of 0.23 seconds. 
Approximately 70 cubes were obtained for each hemisphere dataset, totalling in an integration time of 160 minutes.
A three point dither pattern was used to allow subtraction of the sky background and detector systematics as detailed 
in \citet{Kenworthy13}. Unsaturated short exposure data with the neutral density filter were also taken for photometry.

\section{Creating the Simulated Data Sets}
\label{sec:Method}

Data cubes at each dither position were pairwise 
subtracted to remove the sky background and detector systematics.
The cubes were shifted to move the core PSF into the middle of a square image and bad frames (open loop and poor AO) were discarded 
(5\% of hemisphere 1, 16\% of hemisphere 2, and 7\% of hemisphere 3). 
The three different APP position angle datasets were processed separately. 
Each hemisphere dataset has its own corresponding unsaturated data for photometry. 
    
Fake planets are subsequently used to determine the limiting contrast after image processing.
The unsaturated Fomalhaut data is used to add a fake planet in each saturated frame. 
One fake planet is added at a time, between $0\farcs2$ and $1\farcs0$ in steps of $0\farcs1$ and delta 
magnitudes in steps of 1 mag from dM= 7 to 13. Due to the asymmetric nature of the APP PSF, 
it is also necessary to determine the signal-to-noise of a planet at different position angles. For our analysis we placed a planet 
on opposite sides of the star (PA=45$^{\circ}$ and 225$^{\circ}$ relative to the sky) 
to take into account the asymmetric PSF of the APP. These two PA orientations ensure that the planet is 
on the dark side of the APP in at least two of the hemispheres at once.
The mean of the limiting contrast at each PA is stored.

The final science frames are processed with several different algorithms to 
recover the fake planet signal. All of the algorithms take advantage of the fake planet's 
rotation in the sky around the star to model and subtract the stellar PSF from each image (ADI, \citet{Marois06}). 
Before each algorithm is applied, the innermost region is masked out (r$< 0\farcs15$) where the star has 
saturated the image and no planet could be detected. 
The method of modelling the stellar PSF differs between the algorithms, detailed in the following subsections.

One metric for detectability of planets is signal-to-noise (S/N). It is a measure of the detectability 
of a point source, assuming the noise is Gaussian and decorrelated between diffraction limited elements at that radius.
The equation below is similar to those in the literature, describing local signal-to-noise:

\[\left (\frac{S}{N}  \right )_{planet}= \frac{F_{planet}}{\sigma (r)\sqrt{\pi r_{ap}^{2}} }\]

where $F_{planet}$ is the sum of the planet flux in an aperture with radius $r_{ap}=3$ pixels and $\sigma$ is the root mean 
square of the pixels in a 180$^{\circ}$, 6 pixel wide arc at the same radius, surrounding the star. 

The equation above assumes statistically independent pixels, which in the case of speckle noise limited regimes is 
typically not be the case.  For the sake of consistency with other papers in the literature, we use one of the most 
common definitions of S/N calculation to facilitate comparison with other methods.  This is a widely acknowledged 
issue in this research field, so while the S/N quoted may be off by a scaling factor, the conclusions in 
this paper do not rely on the absolute scaling as we are comparing analysis techniques.

\section{Data Analysis}

\begin{figure}
 \centering
  \epsscale{1.1}
 \plotone{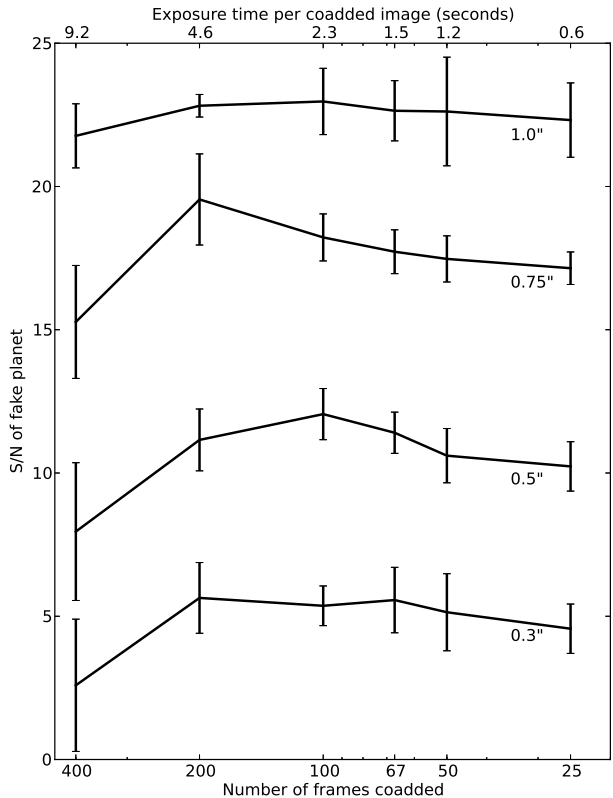}
 \caption{S/N curves for a fake planet at four angular separations, with different amounts of frames coadded. At each angular separation, 
 planets were added at four position angles and averaged. The error bars are 1 sigma. The number of PCs is fixed at 20. Each contrast curve 
 is offset from a S/N of 10 for clarity.
}
 \label{coadds}
\end{figure}

Coadding the frames in a data cube is a common practice but the best number of frames to coadd has not yet been thoroughly studied. 
We experimented with different numbers of coadded frames using fake planets. We ran the data through 
our PCA pipeline (detailed in \ref{sec:PCA}) 
with different numbers of frames coadded (\autoref{coadds}).
For example, 100 frames coadded means that twice as many images are passed to our pipeline as in the 200 coadded frames case. 
\autoref{coadds} shows four S/N curves for planets injected a different angular separations, each with a S/N of approximately 10. The curves are 
offset from 10 for clarity.
Coadding 200 or less frames yields a higher S/N. However, the S/N varies by less than a factor of 2 over all coadds, 
making this a relatively small effect. For the following analysis, we keep the coadds fixed at 200 frames, which yields S/N as good as less coadds, 
but is computationally much faster. This corresponds to 
$\sim$70 coadded images in each hemisphere which are passed to our pipeline. Since there is little field rotation between individual frames 
in a data cube, the smearing effect within a cube is negligible.

\subsection{LOCI}
\label{sec:LOCI}

Locally optimized combination of images (LOCI) \citep{Lafreniere07} is a widely used planet detection 
algorithm which spatially models the stellar PSF to remove speckles. 
An image is divided into rings, which are subdivided into wedges. An optimal, linear combination of 
images subtracts speckles within that region. The least squares fit succeeds at minimizing speckles, 
but also reduces the planet flux through the subtraction for small angular separations.

Each hemisphere dataset is processed with LOCI independently and the final three hemisphere sky aligned cubes are collapsed. 
Since we are using the APP, we only perform LOCI on the ``dark side'' of the image frames. This 180$^{\circ}$ 
D shaped region (inner=${2\lambda}/{D}$, outer=${7\lambda}/{D}$) is the only part of each frame that is coadded in the final image.

\citet{Kenworthy13} analysis of these data used the LOCI algorithm. Monte Carlo simulations exploring LOCI parameters 
ensured that this is the best sensitivity LOCI could produce.

\subsection {Principal Component Analysis}
\label{sec:PCA}

Principal component analysis (PCA) is a mathematical technique that relies on the assumption 
that every image in a stack can be represented as a linear combination of its principal orthogonal components, 
selecting structures that are present in most of the images. Its recent application to high contrast exoplanet 
imaging \citep{Amara12,Soummer12} has been shown to be 
very effective. Unlike LOCI \citep{Lafreniere07} which models the local stellar 
PSF structure, PCA models the global PSF structure. 

The full stack of images with sky rotation is used for PCA. However, since we are using the APP, 
only the ``dark side'' of each image is used in the fit. The S/N from a fake planet is 
lower if we include the ``bright side''. We follow the description of PCA outlined in \citet{Amara12} for the 
following analysis. 

The number of PCs used determines how well the stellar PSF is fit. The first 
few components are the most stable, have less noise, and contain 
the most common structure in all the images. For our default analysis, we used 20 PCs to model the stellar PSF.
PCA is run on each hemisphere dataset independently, as the PCs are correlated with time.
The final de-rotated frames are coadded into one final image covering the full 360$^{\circ}$ around the star. 

The following subsections discuss self-subtraction due to the PCA algorithm as well as 
a series of modifications we performed on PCA to optimize the detection of a 
planet at small $\lambda$/D.

\subsubsection{Self-Subtraction}
\label{sec:self_sub}
Self-subtraction due the LOCI algorithm has been well documented by previous authors \citep{Lafreniere07,Marois10}, but its impact on 
PCA is not yet well studied. The LOCI algorithm requires that the frames nearest in time to the current 
frame are not considered in the least squares fit, thus limiting the self-subtraction of a potential planet. However, 
this frame rejection technique does not completely account for flux loss from a planet.

For our PCA analysis, we draw a distinction between two types of modes: detection and characterization. 
Characterization mode requires fully accounting for flux loss of the planet as a function of number of PCs as we 
map between the measured flux and the calibrated estimate of ``true'' flux. 
However, in detection mode, since we only care about our ability to separate the 
planet signal from the background noise, the main issue is the flux loss relative to the 
separation of the background noise. In this paper, we address simply the detection mode.

\begin{figure}
 \centering 
 \epsscale{1.1}
 \plotone{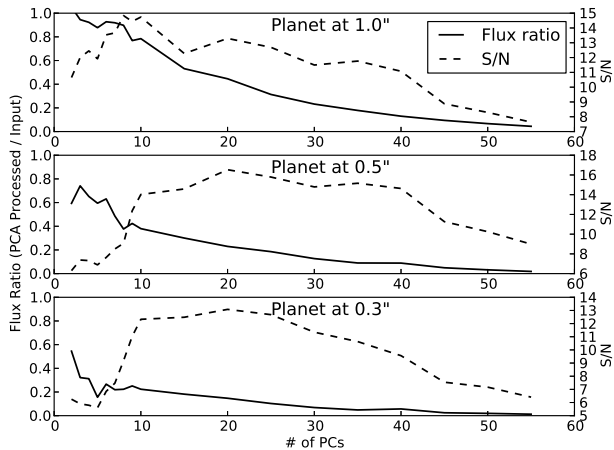}
 \caption{Comparison of the flux ratio and S/N based on number of PCs at different radii. The top figure is for 
 a fake planet at $1\farcs0$, the middle figure is for $0\farcs5$ and the bottom figure is for $0\farcs3$. These 
 figures demonstrate that, while the flux ratio does decrease with PCs, the S/N follows a different curve.}
 \label{PCsvsSN}
\end{figure}

\autoref{PCsvsSN} shows three plots with the flux ratio and S/N vs the number of 
PCs it was processed with. The top figure is for a planet injected at $1\farcs0$, 
middle is at $0\farcs5$ and bottom is at $0\farcs3$. The ``flux ratio'' is the ratio of the injected planet 
flux to the PCA processed flux in a 4 pixel aperture. For each angular separation, the $L'$ 
contrast which yields a S/N of approximately 10 is plotted. This figure demonstrates that 
the PCA method is more efficient at capturing the patterns associated with 
the background fluctuations of the field than capturing information associated with the planet 
translation. This differential effect means that in detection mode, it is acceptable for 
the flux ratio to decrease as long as the noise is decreasing as or more rapidly.

\subsubsection{PCA Modifications}
\label{sec:PCA_mod}

\begin{itemize}[leftmargin=0cm]

\item Frame Rejection

For our PCA code detailed above, all the frames are used in the fit 
and none are rejected. This was done under the assumption that self-subtraction 
of the planet happens less rapidly than the noise subtraction when using PCA. 
As discussed in Section \ref{sec:self_sub}, while we are in ``detection mode'', 
the important factor to consider is the S/N rather than planet flux.
To test this, we used only a subset of the frames to determine the PCs. 
The frames nearest in time to the frame being fitted were rejected. 
These are frames where a potential planet would overlap by 0.5 FWHM or more. 
The number of frames to reject depends on the separation of the planet from the star. The total rotation of the planet is limited by the 
amount of sky rotation achieved during each dataset. A planet very far from its parent star would appear to 
rotate faster between frames, thus less frames need to be rejected. 
This test allows us to compare the S/N of a fake planet processed with standard PCA 
and ``0.5 FWHM Rejection'', where we mimic the routine in LOCI to reject the frames closest in time. 

\item Radius Limited

\begin{figure}
 \centering 
 \epsscale{1.1}
 \plotone{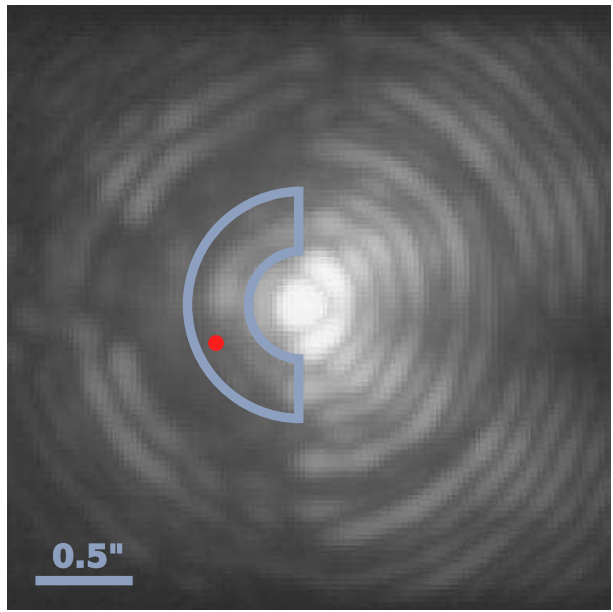}
 \caption{Image demonstrating the APP airy diffraction pattern with the radius limited region
 outlined in blue. This is the only region that is used in the data reduction.}
 \label{app_radius_limited}
\end{figure}

Next, we modified the PCA basis set by only using the image out to a certain radius. 
The outer radius ($R_{out}$) passed to the PCA code determines the amount of information provided to the SVD algorithm. Extra 
information does not necessarily provide a better fit. Our previous applications of PCA kept $R_{out}$ fixed. 
The information passed to the SVD algorithm should be directly related to the stellar PSF. We modified our PCA code to vary 
$R_{out}$ based on the location of the fake planet. The new $R_{out}$ is 1 $\lambda$/D greater than the radius of 
the fake planet (see \autoref{app_radius_limited}), thus performing PCA on a smaller region. 
This experiment was performed to test how significant the stellar PSF fit was affected by 
radii greater than the planet location. 

\item Number of PCs

The main parameter which can be manipulated in PCA is the number of PCs used in the SVD fit. 
The first principal value (the highest singular value in the diagonal matrix) is 
the ``variance'' of the image stack from the mean, in the direction of the first PC. 
The same is true about the second principal value and so on.

\autoref{fom_coeff} shows the PCA coefficient values in descending order for one of our datasets. The first few PCA coefficient values are 
significantly greater than the later values, implying that those PCs contain the most 
dominant features. Increasing the number of PCs in the stellar PSF fit can help bring out the planet signal 
by removing structure, however it also can add noise. Determining the optimal number of 
PCs for a certain stellar PSF fit is an essential but expensive task. The optimal number of 
PCs depends on the time variability of complex speckles.

\begin{figure}
 \centering
  \epsscale{1.1}
 \plotone{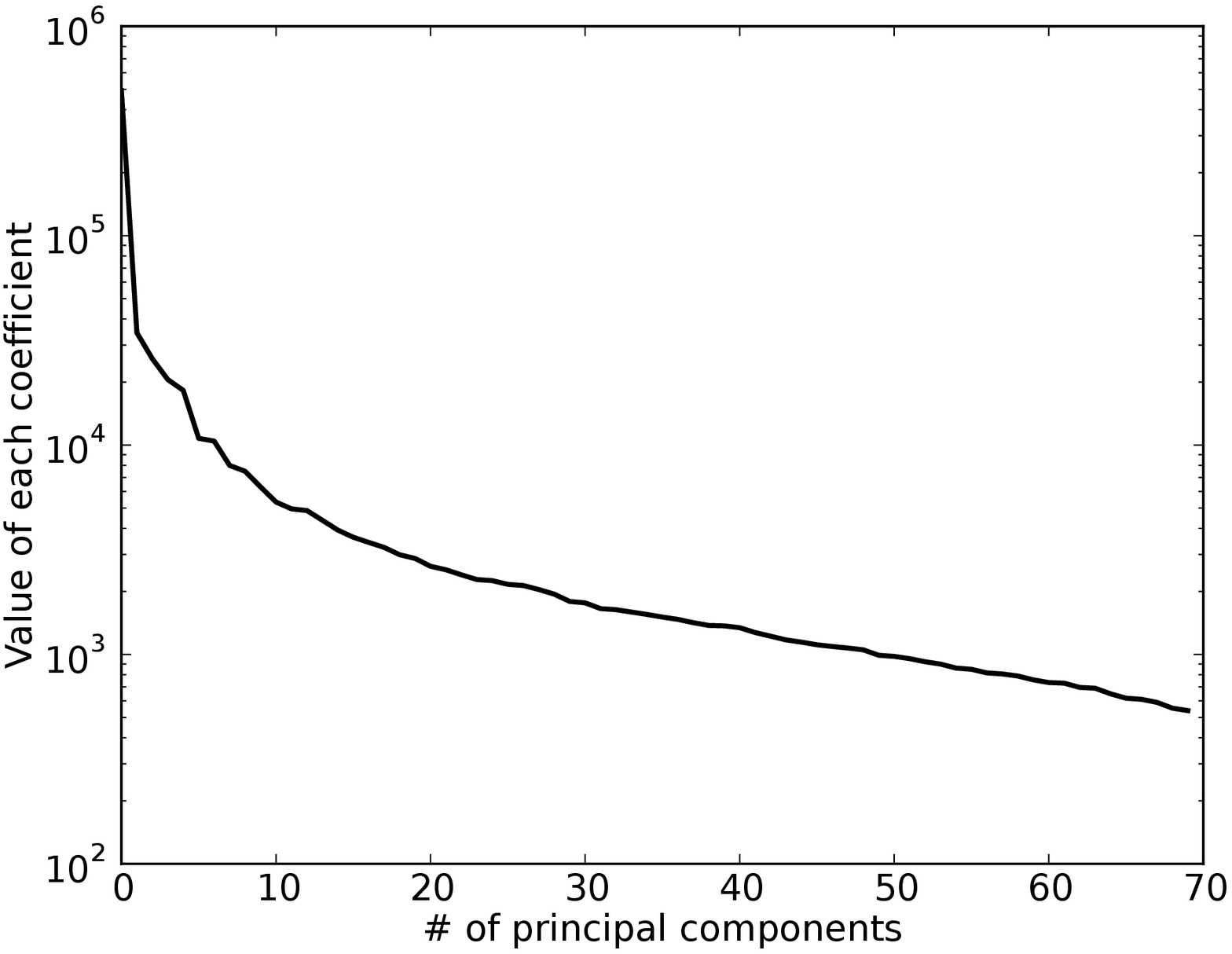}
 \caption{Plot of the PCA coefficient values. The highest PCA coefficient value corresponds to the most 
 significant PC.}
 \label{fom_coeff}
\end{figure}

For each dataset and fake planet angular separation, PCA was run with different numbers of PCs 
ranging from 5 to 60, in increments of 5. 
\end{itemize}

\section{Results and Discussion}

\begin{figure}
 \centering
  \epsscale{1.1}
 \plotone{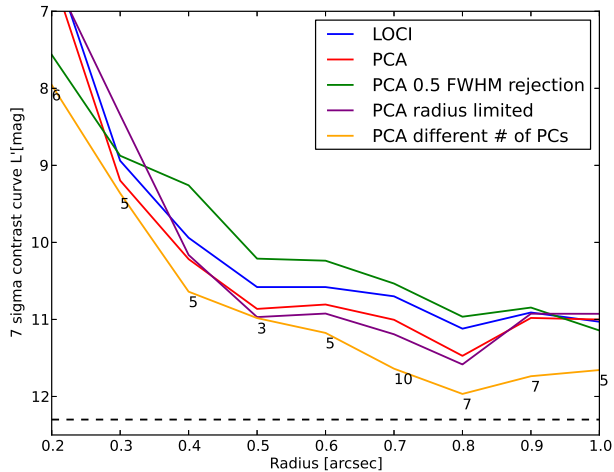}
 \caption{Contrast curves for a 7$\sigma$ detection of a point source in our Fomalhaut APP data 
 processed with LOCI, ADI, and variations of ADI. The LOCI curve is adapted from \citet{Kenworthy13} to a 7$\sigma$ 
 detection. The numbers on the yellow curve signify the number of PCs which yield the highest S/N at that 
 radius. The dashed line is the background limit. The PCA contrast curves are the mean value for fake planets 
 inserted at two PAs on opposite sides of the star (PA=45$^{\circ}$ and 225$^{\circ}$).}
 \label{contrast_7sigma}
\end{figure}

\autoref{contrast_7sigma} shows the results of each image processing method detailed in Section \ref{sec:PCA}. 
Each technique was run with varying planet contrasts at a given radius. We extrapolated between planet contrasts to determine 
contrast that yields a S/N of 7. For the method with varying PCs detailed in Section \ref{sec:PCA_mod}, we noted 
which number of PCs yielded the highest S/N at which radius. These are the numbers listed on the yellow curve in 
\autoref{contrast_7sigma}. 

Our standard PCA technique yields a better contrast curve than LOCI for our 
coronagraphic data. Our modifications to PCA, in some cases, yield better sensitivity. 

Unlike the LOCI algorithm, rejecting the frames nearest in time 
(detailed in Section \ref{sec:PCA_mod}) yields a worse contrast curve than our 
standard PCA. This is likely due to the noise being more correlated in frames closer in time, 
thus providing important information to the SVD algorithm and increasing the 
S/N of the planet. We did not reject any frames in our final data analysis approach.

Limiting the outer radius passed to the SVD algorithm yielded a slightly better contrast ratio than standard PCA from 
$0\farcs5$ to $0\farcs8$. However, this contrast increase is not significant and is only beneficial 
because it is less computationally expensive.

Our standard PCA contrast curve was generated with 20 PCs. By varying the number of PCs we can increase 
the S/N from a companion. Our PC-varying result yields a consistently more sensitive contrast curve then 
all the other methods. We gain between 0.5 and 1 magnitude contrast over our LOCI analysis from $0\farcs2$ to $1\farcs0$. 
From \autoref{contrast_7sigma} we see that the number of PCs which yield the highest S/N for a planet varies based 
on its angular separation.

\begin{figure}
 \centering
  \epsscale{1.1}
 \plotone{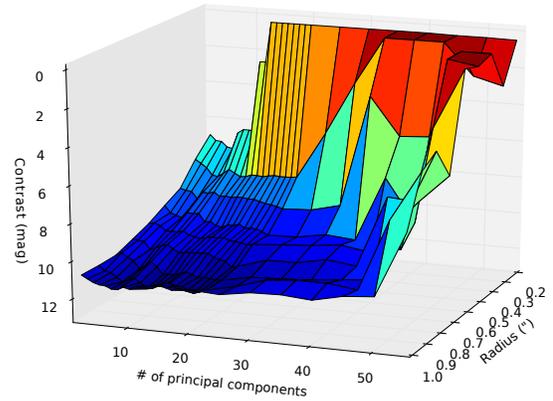}
 \caption{3-D surface of the contrast achieved in a 7$\sigma$ detection with varied numbers of PCs. 
 Varying the number of PCs at small angular separations affects the 7$\sigma$ detection limit by up to 8 magnitudes. Beyond $0\farcs6$, the 
 number of PCs used is less significant.}
 \label{3d_surface}
\end{figure}

\autoref{3d_surface} is a 3-D surface plot showing how the number of PCs at each 
radius affects the contrast at 7$\sigma$ for a planet at a fixed PA. Fake planets were added between 2 and 20 PCs 
in smaller steps to emphasize the structure. This figure demonstrates that 
at small angular separations ($<$ $0\farcs6$), the S/N is sensitive to the number of PCs chosen. 
This is the region where the diffraction and speckles due to the star are more significant than the unstructured noise from thermal 
emission and the sky background. For example, at $0\farcs2$ choosing a small number of PCs yields an 8 magnitude gain in sensitivity 
than a large number of PCs. Increasing the number of PCs 
quickly leads to nearly complete self-subtraction. This can be seen in \autoref{3d_surface} as a contrast of nearly zero. 
As we move to larger radii the optimal number of PCs remains in the 5-20 PC range. Beyond $0\farcs6$
where the number of PCs shows no significant preference below 45 PCs. 

\subsection{Comparison with Kenworthy et al. (2013)}

\begin{figure}
 \centering
  \epsscale{1.1}
 \plotone{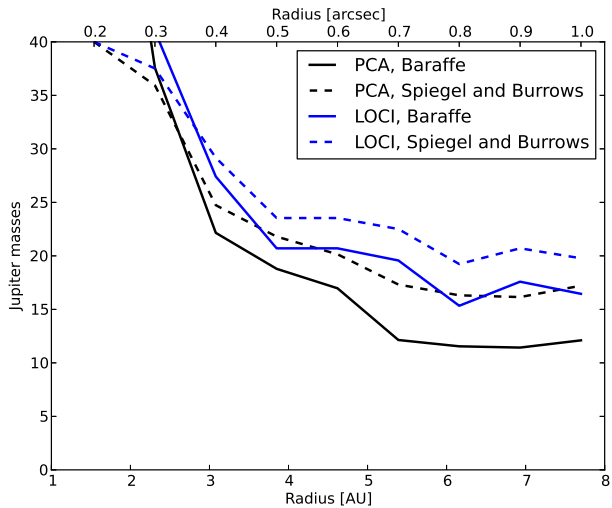}
 \caption{Detection limit for fake companions around Fomalhaut generated with PCA (black lines) and LOCI (blue lines, converted to 7$\sigma$ detection 
 from \citet{Kenworthy13}) using \citet{Baraffe03} (solid lines) and \citet{Spiegel12} (dashed lines). }
 \label{mass_limit}
\end{figure}

Our PCA re-analysis of these data improves sensitivity at small inner working angles, from $0\farcs2$ to $1\arcsec$,
in some cases by 1 magnitude (see blue and yellow curves, \autoref{contrast_7sigma}). We convert the best 7$\sigma$ detection 
contrast curve to an upper mass limit for planets using the \citet{Baraffe03} and \citet{Spiegel12} atmospheric models (\autoref{mass_limit}) 
assuming an age of 440 Myr \citep{Mamajek12}. 
We confirm the non-detection of companions with a model-dependent upper mass limit of 13-18 M$_{\text{Jup}}$ from 4-10 AU. Our new 
upper mass limit is based on our more robust 7$\sigma$ detection limit. The 1 magnitude increase in the contrast ratio 
at $0\farcs5$ translates to an increased sensitivity of $\Delta$ 7 $M_{\text{Jup}}$. 
The increase in sensitivity allows us to probe planetary masses ($<$15 M$_{\text{Jup}}$)
at small angular separations. 

\subsection{Fainter Fomalhaut}

We have shown that the number of PCs which yield the highest signal-to-noise depends on 
the planet's distance from the parent star (yellow line, \autoref{contrast_7sigma}). 
At small angular separations ($<$ $0\farcs6$), the S/N is sensitive to the number of PCs chosen (\autoref{3d_surface}). 
This is the limit where the diffraction from the central star is equal to or less 
significant than the background noise.

We add Gaussian white noise to our data to test if this turnover point changes for a fainter target. 
Increasing the sky background noise makes Fomalhaut 1.5 magnitudes fainter, while 
keeping the telescope conditions and Strehl identical. This is the ideal way to test 
how fainter targets will behave. Fake planets are once again injected and the best number 
of PCs at each angular separation is noted. 

\begin{figure}
 \centering
  \epsscale{1.1}
 \plotone{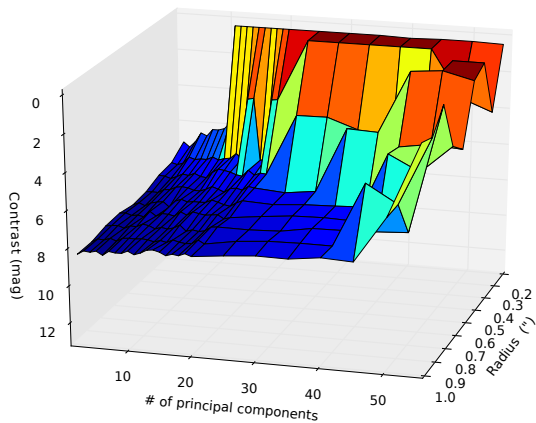}
 \caption{Similar to the 3-D surface in \autoref{3d_surface}, but with Gaussian white noise added to the data. The resulting 
 star is 1.5 magnitudes fainter than Fomalhaut. 
}
 \label{3d_surface_noise}
\end{figure}

Changing the number of PCs used at each angular separation is still the best method for detecting companions. 
As expected, the regime of large numbers of PCs at small separations results in low contrast, which then improves down to a 
plateau at smaller PCs and larger radii. The turnover point remains near $0\farcs6$, beyond which the diffraction 
from the star is no longer significant and the optimal number of PCs is less clear. Beyond this 
separation, the background noise dominates the SVD fit and thus does not help subtract the stellar PSF.

\section{Conclusion}

We re-analyze our Fomalhaut APP/NaCo/NB4.05 data using PCA and compare it with the 
LOCI algorithm. PCA yields a more sensitive contrast curve than the LOCI algorithm at small inner working angles. 
We tested several modifications to PCA and gain up to 1 mag of contrast over our LOCI analysis from $0\farcs2$ to $1\farcs0$. 
The most effective parameter which optimized PCA was varying the number of principal components. 
The number of principal components chosen is sensitive for planets at small inner working angles. The detection limit of a planet 
at small radii can vary by several magnitudes. Careful attention should be paid to determining the number of principal components 
used at radii where the speckles are more significant than the unstructured noise of thermal emission and the sky background.
Running PCA for a range of principal components at each angular separation and generating a 3-D surface is a useful way 
to visualize the optimal number of principal components needed to pull out a faint planet signal. 

Futher analysis is needed in other wavelengths, as differing Strehl ratios may affect the turnover point 
where the stellar diffraction is less significant than the background noise. 
These results have direct application for current and future planet imaging campaigns, which will likely use 
a combination of PCA, LOCI, and other image processing techniques.

\bibliographystyle{apj}

\end{document}